\documentclass{ckm}                 

\usepackage{txfonts}            
\usepackage{relsize}
\def\CP{$ C \! P$ }
\def\babar{\mbox{\slshape B\kern-0.1em{\smaller A}\kern-0.1em
    B\kern-0.1em{\smaller A\kern-0.2em R}}}
\confname{Workshop on the CKM Unitarity Triangle, IPPP Durham, April
  2003}

\title{Radiative Penguin Decays from \babar}

\author{G Eigen (representing the \babar \ collaboration)}
\address{Department of Physics, University of Bergen}

\begin{document}
\begin{abstract}
We summarize the latest \babar \ results on $B \rightarrow K^{(*)} \ell^+ \ell^-$
and $B \rightarrow  \rho(\omega) \gamma$.
\end{abstract}

\maketitle


\section{Introduction}

Electroweak penguin decays provide a promising hunting ground for Physics 
beyond the Standard Model (SM). The decay $B \rightarrow X_s \gamma$, which
proceeds through an electromagnetic penguin loop, already provides stringent
constraints on the supersymmetric (SUSY) parameter space \cite{ali}. 
The present data samples of $\sim 1 \times 10^8 B \bar B$ events allow to
explore radiative penguin decays with branching fractions of the order of 
$10^{-6}$ or less. In this brief report we discuss a study of
$B \rightarrow K^{(*)} \ell^+ \ell^-$ decay modes
and a search for $B \rightarrow \rho (\omega) \gamma$ decays.

\section{Study of $B \rightarrow K\ell^+ \ell^-$ and 
$B\rightarrow K^* \ell^+ \ell^-$}

\begin{figure}
\hbox to\hsize{\hss
\includegraphics[width=\hsize]{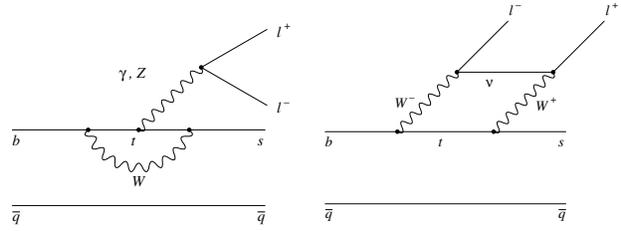}
\hss}
\caption{Lowest-order diagrams for $B \rightarrow K^{(*)} \ell^+ \ell^-$ in 
SM.}
\label{fig:kll-fd}
\end{figure}

\begin{figure}
\hbox to\hsize{\hss
\includegraphics[width=\hsize]{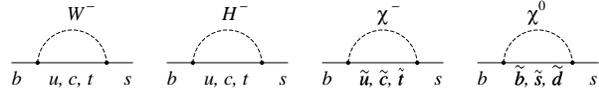}
\hss}
\caption{Penguin loop diagrams involving a $W$ boson, a charged Higgs boson,
a chargino and a neutralino, respectively.} 
\label{fig:kll-np}
\end{figure}

The decays $B \rightarrow K \ell^+ \ell^-$ and 
$B \rightarrow K^* \ell^+ \ell^-$ proceed through an electromagnetic 
penguin loop, a $Z^0$ penguin loop or a weak box diagram as shown 
in Figure~\ref{fig:kll-fd}. In the framework of the operator 
product expansion (OPE) the decay rate is 
factorized into perturbatively calculable short-distance contributions that 
are parameterized by scale-dependent Wilson coefficients and non-perturbative 
long-distance effects that are represented by local four-quark operators. 
Operator mixing occurring in next-to-leading order perturbation theory leads 
to three effective scale-dependent ($\mu_b$)
Wilson coefficients, $C_7^{eff}(\mu_b),
C_9^{eff}(\mu_b)$, and $C_{10}^{eff}(\mu_b)$ that are each sensitive to
New Physics contributions. Examples for non-SM penguin loops are depicted
in Figure~\ref{fig:kll-np}.
In the Standard Model the branching fractions are
predicted to be within the following ranges, 
${\cal B}(B \rightarrow K \ell^+ \ell^-) =(0.23-0.97) \times 10^{-6}$,
${\cal B}(B \rightarrow K^* \mu^+ \mu^-) =(0.81-2.64) \times 10^{-6}$ and 
${\cal B}(B \rightarrow K^* e^+ e^-) =(1.09-3.0) \times 10^{-6}$ 
\cite{kll-th}. 
In supersymmetric models, for example,
the branching fractions my be enhanced by more than a factor of two \cite{ali}.

\begin{figure}
\hbox to\hsize{\hss
\includegraphics[width=\hsize]{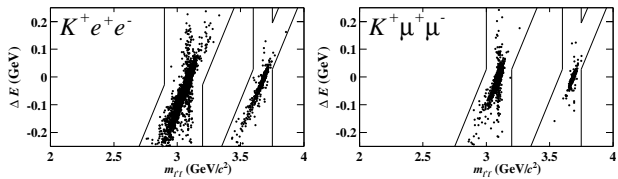}
\hss}
\caption{Charmonium veto (hatched regions) in the 
$\Delta E^*- m_{\ell^+ \ell^-}$ plane for 
$B \rightarrow K^{(*)} e^+ e^-$ and 
$B \rightarrow K^{(*)} \mu^+ \mu^-$. 
The dots represent
simulations for $B \rightarrow J/\psi 
( \rightarrow \ell^+ \ell^-) K^{(*)}$ and 
$B \rightarrow \psi(2S) ( \rightarrow \ell^+ \ell^-) K^{(*)}$.}
\label{fig:ch-veto}
\end{figure}

\babar \ has analyzed eight final states where a $
K^\pm, K^0_S, K^{*0}$ or $K^{*\pm}$ recoils
against a $\mu^+ \mu^-$ or $e^+ e^-$ pair, using an integrated luminosity of 
$ 77.8\ \rm fb^{-1}$ that corresponds to $(84.4\pm 0.9) \times 10^6~B \bar B$ 
events. 
The discriminating variables are beam energy-substituted 
mass \break
$m_{ES}=\sqrt{(E_{beam}^*)^2 -(\vec p_B^*)^2}$  and the energy difference 
\break
$\Delta E^*=E^*_B -E^*_{beam}$, where  $\vec p_B$, $ E^*_B$ and $E^*_{beam}$ 
denote the B-momentum, B-energy and beam energy in the center-of-mass 
(CM) frame, respectively. The 
$\Delta E^*-m_{ES} plane$ is divided into three regions, a signal region 
($\pm 3 \sigma$ boxes around the signal), the fit region 
($m_{ES} > 5.2~\rm GeV$, \break
$\mid \Delta E^* \mid < 250~\rm MeV$), and a large 
side band  ($m_{ES} > 5.0~\rm GeV$, $\mid \Delta E^* \mid < 500~\rm MeV$).
Specific selection criteria are used to suppress individual backgrounds. 
A Fisher discriminant \cite{fisher} is used to eliminate the 
back-to-back continuum background. It is based on event shape variables
such as the thrust angle 
between the daughter particles of the $B$ meson candidate and that of the 
remaining particles in the event ($\cos \theta_T$), the B decay angle in the 
$\Upsilon(4S)$ rest frame between the B candidate and the beam axis
($\cos \theta^*_B$), and
the ratio of second-to-zeroth Fox-Wolfram moments ($R_2$) \cite{FW}, 
as well as the invariant 
mass of the $K-\ell$ system $m_{k\ell}$. 
To discriminate against combinatorial $B \bar B$ background 
a likelihood function is used that combines the missing energy in the event
($E_{miss}$), with the dilepton vertex probability, the significance of the 
dilepton separation along the beam direction and $\cos \theta^*_B$.
To reject events from $B \rightarrow J/\psi K^{(*)}$ and 
$B \rightarrow \psi(2S) K^{(*)}$ decays with 
$J/\psi (\psi(2S) \rightarrow \ell^+\ell^-$
that have the same event topologies in a restricted $m_{\ell\ell}$ mass region
as signal events, the shaded regions in the $\Delta E^*-
m_{\ell \ell}$ plane shown in Figure~\ref{fig:ch-veto} are vetoed. 
The inclined bands provide an efficient 
rejection of $J/\psi K^{(*)}$ events in the fit region, in which one or both 
leptons radiated a photon.

The selection criteria are optimized on simulated data as well as data 
sidebands. The simulated data are used to determine efficiencies and 
estimate the contribution of backgrounds that may peak in the 
$\Delta E^* - m_{ES}$ signal box.
Specifically, the decay channels
$B \rightarrow K^* \gamma$,
$B \rightarrow J/\psi K^{(*)}$, $B \rightarrow \psi(2S) K^{(*)}$,
$B \rightarrow K^{(*)} \pi^0$, $B \rightarrow K^{(*)} \eta$,
$B \rightarrow K \pi \pi$, $B \rightarrow K K \pi$, $B \rightarrow K K K$,
and $B \rightarrow D\pi $ have been studied.
The Monte Carlo results are checked with data 
control samples, including exclusive and inclusive charmonium decays, 
$B \rightarrow D \pi$ decays, data sidebands and $K^{(*)} e^\pm \mu^\mp$ 
samples. For the combined exclusive charmonium modes we achieve
an excellent agreement between data and Monte Carlo for the ratio of 
event yields of $1.013 \pm 0.018$.
Further details of the event selection are discussed in 
\cite{kll}. 

In each of the four $K \ell^+ \ell^-$ final states,
a signal is extracted from a two-dimensional fit to the $m_{ES}-\Delta E^*$ 
plane. Similarly, a signal yield for the four $K^* \ell^+ \ell^-$ final states
is obtained from a three-dimensional fit to the $m_{ES}-\Delta E^*-m_{k \pi}$
space. The signal shapes are 
obtained from Monte Carlo samples with fine-tuning on the exclusive 
charmonium modes. To account for the asymmetric $\Delta E^*$ and $M_{ES}$ 
distributions that result from radiation effects of the final-state 
leptons and correlations in the two variables a product of Crystal Ball 
functions is used \cite{xtal}. The combinatorial backgrounds are 
parameterized with ARGUS functions \cite{argus}, where both the normalization 
and the shape parameters are left free. Only in $B \rightarrow K^\pm e^+ e^-$,
a significant yield of $16.2 ^{+4.8}_{-4.1}$ events is observed. The selection 
efficiency is $(19.1\pm 1.1)\%$, yielding a branching fraction of 
\begin{equation}
{\cal B}(B^+ \rightarrow K^+ e^+ e^-)=(0.96^{+0.28}_{-0.24}\pm 0.06)\times 
10^{-6}
\end{equation}
The second largest yield of $(7.8^{+5.4}_{-4.2})$ events
is found for $B^0 \rightarrow K^{*0} e^+ e^-$ for which a selection
efficiency of $(10.6\pm 0.9)\%$ is achieved. The individual results 
for each of the eight final states are summarized in Table~\ref{tab:kll}. 

\begin{figure}
\hbox to\hsize{\hss
\includegraphics[width=\hsize]{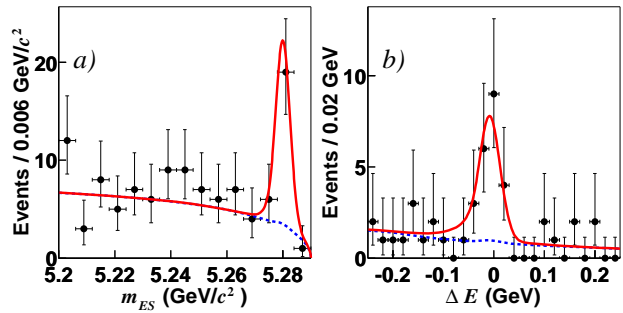}
\hss}
\caption{Projections of the two-dimensional fit of the combined
$K \ell^+ \ell^-$ data sample for a)
$m_{ES}$ after the constraint $\rm -0.11 < \Delta E^* < 0.05~GeV$ and b)
$\Delta E^*$ after requiring $|m_{ES} - m_B | \rm < 6.6~MeV/c^2$.
The solid and dashed curves show the entire
fit and the total background contribution, respectively.}
\label{fig:kll-sum1}
\end{figure}

\begin{figure}
\hbox to\hsize{\hss
\includegraphics[width=\hsize]{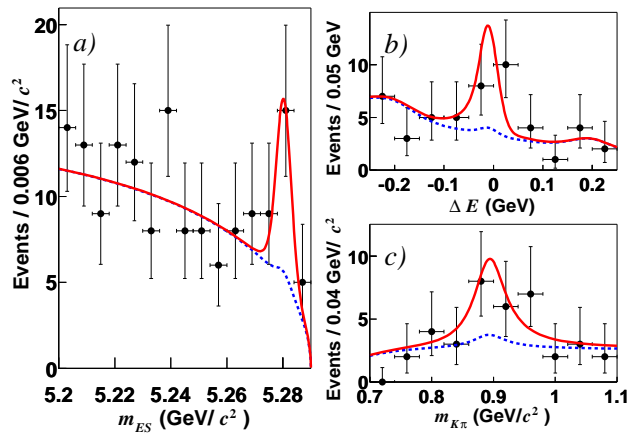}
\hss}
\caption{Projections of the tree-dimensional fit of the combined
$K^* \ell^+ \ell^-$ data sample for a)
$m_{ES}$ after the constraint $\rm -0.11 < \Delta E^* < 0.05~GeV$ and
$0.817 < m_{K\pi} < 0.967~MeV$, b)
$\Delta E^*$ after requiring $|m_{ES} - m_B | \rm < 6.6~MeV/c^2$ and
$0.817 < m_{K\pi} < 0.967~MeV$, and c) 
$m_{K \pi}$ satisfying $\rm -0.11 < \Delta E^* < 0.05~GeV$
and  $|m_{ES} - m_B | \rm < 6.6~MeV/c^2$.
The solid and dashed curves show the entire
fit and the total background contribution, respectively.}
\label{fig:kll-sum2}
\end{figure}

The $M_{ES}$ and $\Delta E^*$ projections of the two-dimensional fit for 
the combined $B \rightarrow K \ell^+ \ell^-$ final states
are shown in Figure~\ref{fig:kll-sum1}. The corresponding distributions 
of the three-dimensional fit for the combined
$B \rightarrow K^* \ell^+ \ell^-$ channels as well as the $m_{K\pi}$ mass 
projection are shown in Figure~\ref{fig:kll-sum2}. In order to combine 
$K^* \mu^+ \mu^-$ and $K^* e^+ e^-$ results their ratio of branching 
fractions is assumed to be 
${\cal B}(B \rightarrow K^* e^+ e^-)/{\cal B}(B \rightarrow K^* \mu^+ \mu^-)
=1.33$ \cite{ali}. The multiplicative systematic errors that do
not affect the significance range from $6\%-12\%$ 
for $K \ell^+ \ell^-$ modes and $7\%-15\%$ for $K^* \ell^+ \ell^-$ modes. 
The largest contributions result from the $K^0_S$ efficiency ($10\%$), the 
model dependence ($4\%-7\%$), $\mu$ identification ($3.2\%$), 
$B \bar B$ likelihood ratio ($1.3\%-5.6\%$), the Fisher discriminant
($ 0.9\%-3.1\%$), and the tracking efficiency for hadrons
($1.3\%-3.9\%$).
Contributions from $K/\pi$ identification, $e$ identification, tracking
efficiency for leptons, Monte Carlo statistics and 
$B \bar B$ counting are small ($1\%-2\%$). The additive systematic 
errors that affect the significance
result from the signal yields in the fit, including uncertainties in 
signal shapes, background shapes and the amount of peaking backgrounds. 
Including systematic errors the significance of the combined 
$K \ell^+ \ell^-$ sample is $7.0\sigma$, while that of the combined 
$K^* \ell^+ \ell^-$ sample is $3.0\sigma$.

For the combined 
$B \rightarrow K \ell^+ \ell^-$ modes we measure a branching fraction of

\begin{equation}
{\cal B}(B \rightarrow K e^+ e^-)=(0.68^{+0.17}_{-0.15}\pm 0.04)\times 
10^{-6}.
\end{equation}
For the combined  
$B \rightarrow K^* \ell^+ \ell^-$ modes we determine a branching fraction 
of
\begin{equation}
{\cal B}(B^+ \rightarrow K^*e^+ e^-)=(1.4^{+0.57}_{-0.49}\pm 0.21 
\times 10^{-6}.
\end{equation}
The \babar \ results are consistent with those obtained by BELLE \cite{belle} 
and most SM predictions, but ${\cal B}(B \rightarrow K \ell^+ \ell^-)$ 
is almost $2 \sigma$ higher than a recent next-to-next-to-leading 
order prediction of 
${\cal B}(B^+ \rightarrow K \ell^+ \ell^-)= (0.35 \pm 0.12) \times 10^{-6} $ 
 \cite{ali}. 

\begin{table}
\caption{Measured event yields, efficiencies and branching fractions for 
$B \rightarrow K \ell^+ \ell^-$ and $B \rightarrow K^* \ell^+ \ell^-$ modes.
\break }
\label{tab:kll}
\begin{tabular}{|l|l|l|l|}
\hline
$ \ \ \ \ \ \ \ $ Mode & $\ \ $yield & $\ \ \ \ \epsilon [\%]$ & 
$\ \ \ \ \ {\cal B}\times 10^6$ \\
 & [events] &  &  \\
\hline
$B^+ \rightarrow K^+ e^+ e^-$ & $16.2^{+4.8}_{-4.1}$& $19.1\pm 1.1$& $0.96
^{+0.28+0.06}_{-0.24-0.06}$ \\
$B^+ \rightarrow K^+ \mu^+ \mu^-$ & $1.0^{+2.0}_{-1.2}$ &$8.7\pm 0.6$& $0.14
^{+0.25+0.02}_{-0.16-0.02}$ \\
$B^0 \rightarrow K^0 e^+ e^-$ & $-0.5^{+1.6}_{-1.0} $& $20.6\pm 2.5$& $-0.08
^{+0.25+0.05}_{-0.16-0.05} $\\
$B^0 \rightarrow K^0 \mu^+ \mu^-$ & $5.0^{+2.8}_{-2.0} $&$9.2\pm 0.7$ & $1.78
^{+0.99+0.22}_{-0.73-0.22} $ \\
$B^0 \rightarrow K^{*0} e^+ e^-$ & $7.8^{+5.4}_{-4.2} $ &$13.5 \pm 0.9$& 
$0.98^{+0.68+0.18}_{-0.54-0.18}$ \\
$B^0 \rightarrow K^{*0} \mu^+ \mu^-$ & $4.5^{+3.8}_{-2.8}$ & $6.5 \pm 0.6$& 
$1.18^{+0.99+0.26}_{-0.73-0.26} $ \\
$B^+ \rightarrow K^{*+} e^+ e^-$ & $2.7^{+4.1}_{-2.9}$ & $10.2\pm 1.4$ & $1.31
^{+1.97+0.33}_{-1.38-0.33} $ \\
$B^+ \rightarrow K^{*+} \mu^+ \mu^-$ & $4.4^{+3.5}_{-2.4} $ &$ 5.0 \pm 0.8$& 
$4.33^{+3.41+0.68}_{-2.40-0.68}  $ \\
\hline 
\end{tabular}
\end{table}

\section{Search for $B \rightarrow \rho (\omega) \gamma$}

The decays $B \rightarrow \rho (\omega) \gamma$ are flavor-changing neutral 
current $b \rightarrow d$ transitions that are mediated by an electromagnetic 
penguin loop. The decay rates depend on the effective 
Wilson coefficient $C^{eff}_7(\mu_b)$ that may be enhanced by New Physics 
contributions as in $B \rightarrow K^* \gamma$. 
They are, however, suppressed with respect to $B \rightarrow K^* \gamma$ 
by $|V_{td}/V_{ts}| ^2$. Thus, measuring the ratio of branching fractions 
${\cal B}(B \rightarrow \rho \gamma)/{\cal B}(B \rightarrow K^* \gamma)$,
allows us to extract $|V_{td}/V_{ts}|$. In SM the branching fraction for the 
charged mode in next-to-leading order is predicted to lie between 
${\cal B}(B^+ \rightarrow \rho^+ \gamma)=(0.85\pm0.4)\times 10^{-6}$ 
\cite{ali2} and ${\cal B}(B\rightarrow \rho^+ \gamma)=(1.58^{+0.53}_{-0.46}) 
\times 10^{-6}$ \cite{bosch}.
For the neutral modes the branching fractions are a factor of two smaller 
than that for $B^+ \rightarrow \rho^+ \gamma$ due to isospin. 
By determining the ratio of decay rates
$\Gamma(B \rightarrow \rho \gamma)/\Gamma(B \rightarrow K^* \gamma)$
uncertainties due to scale dependence and a precise knowledge of 
form factors cancel except for $\rm SU(3)$ breaking effects.
The largest remaining theoretical uncertainties 
result from contributions of $W$-annihilation/W-exchange diagrams, 
which amount to $\sim 20\%$ in $B^+ \rightarrow \rho^+ \gamma$ and to 
$\sim 13\%$ in $B^0 \rightarrow \rho^0 (\omega) \gamma$ \cite{ali2}.

\babar \ has searched for $B \rightarrow \rho (\omega) \gamma$ modes using an 
integrated luminosity of $77.8~\rm fb^{-1} $ on the $\Upsilon (4S)$ peak and 
$9.6~\rm fb^{-1}$ in the continuum 40~MeV below the $\Upsilon(4S)$ peak. 
Challenges in the analysis stem from a huge $q \bar q$ continuum background 
including initial-state radiation, background from $B \rightarrow K^* \gamma$ 
and the fact that the $\rho$ resonance is much broader than the $K^*$ 
resonance. The $q \bar q \gamma$ continuum background with a hard 
$\gamma$ from initial-state radiation may have a similar event shape as that 
of the signal which is less spherical than a typical $B \bar B$ 
event. Photon candidates with energies of 
$\rm 1.5~GeV < E_\gamma < 3.5~GeV$ that are inconsistent with originating 
from a $\pi^0$ or $\eta$ decay are combined with a 
$\rho^+, \rho^0$, or $\omega$ candidate, where the vector mesons
are reconstructed from $\pi^+ \pi^0$, $\pi^+ \pi^-$ and three-pion 
combinations, respectively. 
Rejecting charged tracks in the signal that are 
consistent with a kaon, the $K/\pi$ misidentification is less than $1\%$. 
The $\pi \pi$ ($3\pi$) invariant mass has to lie within a mass window of
$\rm 520-1020~MeV/c^2 \  (759.6-805.6~MeV/c^2)$ and its momentum 
in the CM frame must satisfy $\rm 2.3 < p^*_{\pi\pi} < 2.85~ GeV/c$
($\rm 2.4 < p^*_{3\pi} < 2.8~ GeV/c$).
A $\pi^0$ candidate must have a $\gamma \gamma$ invariant 
mass of $\rm 115 < m_{\gamma \gamma} < 150~MeV/c^2$. To improve the
momentum resolution we perform a kinematic fit with $m_{\gamma \gamma}$ 
constrained to the nominal $\pi^0$ mass.

To reduce the $q \bar q$ continuum background a neural network is used that 
is based on event-shape variables ($\cos \theta_{thrust}, \cos \theta^*_B, 
\cos \theta_{helicity}$), the Dalitz decay angle for $\omega$, the energy flow 
in 18 cones around photon direction, the vertex separation $\Delta z$, the 
ratio of second-to-zeroth Fox Wolfram moment, $R_2^\prime$, calculated in a 
frame recoiling the photon and the net flavor in the event \cite{babar5}.
The neural network is trained with Monte Carlo 
signal events and continuum data. The neural network output is cross-checked 
with data using a $B^0 \rightarrow D^- \pi^+$ sample
that has a similar topology as 
the signal and the off-resonance data sample. Further details of the 
event selection are discussed in \cite{rhog}.

\begin{figure}
\hbox to\hsize{\hss
\includegraphics[width=\hsize]{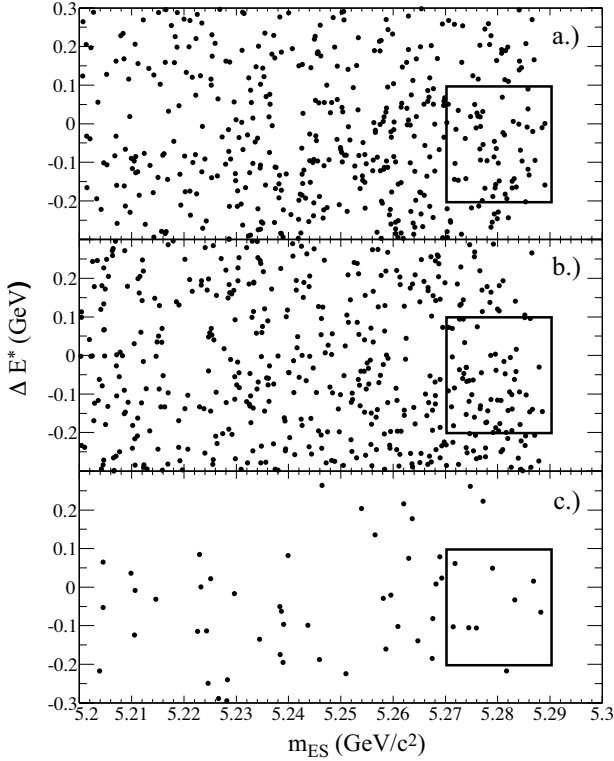}
\hss}
\caption{$\Delta E^* - m_{ES}$ scatter plots of the fit region for a) 
$B^+ \rightarrow \rho^+ \gamma$, b) $B^0 \rightarrow \rho^0 \gamma$ and c)
$B^0 \rightarrow \omega \gamma$ candidates. The boxes indicate the expected
signal regions}
\label{fig:rhog}
\end{figure}

The $\Delta E^* - m_{ES}$ distributions for the final data samples are
shown in Figure~\ref{fig:rhog}. The signal yields are extracted
from a maximum likelihood fit in the three-dimensional space 
$\Delta E^* - m_{ES} - m_\rho(m_\omega)$. 
The procedure is crosschecked 
with our $B \rightarrow K^* \gamma $ data samples. For 
$B \rightarrow K^{*0} \ (K^{*+} )\gamma$ the maximum likelihood
fit yields $343.2 \pm 21.0\ 
(93.1\pm 12.6)$ events, which is in good agreement with the
expected yields of $332\pm 36 \ (105 \pm 18)$ events, respectively. 

The extracted signal yields of $4.8^{+5.2}_{-4.7}$ events for 
$B \rightarrow \rho^0\gamma$, 
$6.2^{+7.2}_{-6.2}$ events for $B \rightarrow \rho^+\gamma$, 
and $0.1^{+2.7}_{-2.0}$ events for $B \rightarrow \omega \gamma$ 
are consistent with background. The  
efficiencies are $12.3\%$, $9.2\%$ and $4.6\%$, respectively. Including 
systematic errors, which respectively increase from $11.8\%$ to $13.4\%$ and 
$17.3\%$ for the three decay channels,  
we obtain branching fraction upper limits $ @\ 90\% \ CL$  
of ${\cal B} (B^0 \rightarrow \rho^0\gamma) < 1.2 \times 10^{-6}, \hfill \break
{\cal B} (B^+ \rightarrow \rho^+\gamma) < 2.1 \times 10^{-6}$, 
and
${\cal B} (B^0 \rightarrow \omega\gamma) < 1.0 \times 10^{-6}$.
These limits are significantly lower than those of previous searches
\cite{cleo2}\cite{belle2}. 

Assuming isospin symmetry the $\rho^+ \gamma$ and $\rho^0 \gamma$ samples 
are combined yielding 
${\cal B} (B \rightarrow \rho\gamma) < 1.9 \times 10^{-6}$. Using the recent
\babar \ $B \rightarrow K^* \gamma$ branching fraction measurement \cite{babar3} 
this translates into an upper limit on the ratio of branching fractions of
${\cal B}(B \rightarrow \rho \gamma)/ {\cal B}(B \rightarrow K^* \gamma) < 
0.047$ $@ \ 90\% \ CL$. To constrain the ratio of $|V_{td}/V_{ts}|$ we use
the parameterization \cite{ali2}.

\begin{equation}
\frac{{\cal B}(B \rightarrow \rho \gamma)}{{\cal B}(B \rightarrow K^* \gamma})
 = \Biggl| \frac{V_{td}}{V_{ts}}  \Biggr|^2 
\Biggl( \frac{1 - m^2_\rho/M^2_B}{1 - m^2_{K^*}/M^2_B} \Biggr)^3 
\zeta^2[1+\Delta R].
\end{equation}
\noindent
The parameter $\zeta$ represents $SU(3)$ breaking while $\Delta R$ accounts for
the annihilation diagram in $B^+ \rightarrow \rho^+ \gamma$. Using
$\zeta=0.76\pm0.1$ and $\Delta R=0.0\pm 0.2$ (\cite{ali2},  \cite{GP})
we obtain an upper limit of 
$|V_{td}/V_{ts}| < 0.34 \ @ \ 90\% \ CL$. This is still larger than the limit 
of $|V_{td}/V_{ts}| < 0.23 \ @ \ 95\% \ CL$ which is derived from
$B_s \bar B_s$ and $B_d \bar B_d$ mixing results 
for $\Delta m_{B_d} =0.503 \pm 0.006\ \rm 
ps^{-1}$, $\Delta m_{B_s} > 14.4 \ \rm ps^{-1} \ @ \ 95\% \ CL$ and 
$\xi =1.24$ \cite{ckmws}.

The present upper limits are approaching the theoretical predictions. Assuming
a branching fraction of ${\cal B}(B^+ \rightarrow \rho^+ \gamma) =
1 \times 10^{-6}$ and ${\cal B}(B^0 \rightarrow \rho^0 \gamma =
{\cal B}(B \rightarrow \omega \gamma = \frac{1}{2}
{\cal B}(B^+ \rightarrow \rho^+ \gamma)$ 
we estimate a signal significance and experimental errors 
for different luminosities as listed in 
Table~\ref{tab:rhog}. The small values for the significance 
($i.e.$ large experimental errors on the branching 
fraction and on  $|V_{td}/V_{ts}|$) result from the present event selection,
while the large (small) values are obtained by assuming that for the same
selection efficiency as in the present analysis the background is
halved. For a luminosity 
of $\sim 500 \ fb^{-1}$ a significant measurement of these modes is expected. 
The limiting factor for extracting $|V_{td}/V_{ts}|$ is 
the theoretical
uncertainty from $SU(3)$ breaking and the size of the annihilation diagram. 
The latter uncertainty can be removed by using only 
$B^0 \rightarrow \rho^0 \gamma$ events, but for the same significance
a factor of four increase in luminosity is required.

\begin{table}
\caption{Extrapolations of the
significance, experimental error on the branching fraction
and experimental error on $|V_{td}/V_{ts}|$ for the combined $B \rightarrow 
\rho (\omega) \gamma $ modes expected for different luminosities.
\break }
\label{tab:rhog}
\begin{tabular}{|l|l|l|l|}
\hline
Luminosity & significance & $(\sigma_{\cal B}/{\cal B})_{exp}$ &
$\sigma(V_{td}/V_{ts})$ \\
\hline
$100\ \rm fb^{-1}$ & $1.9-2.8 \ \sigma$ & 0.38-0.53 &0.19-0.27 \\ 
$200\ \rm fb^{-1}$ & $2.7-3.9 \ \sigma$ & 0.28-0.38 &0.14-0.19 \\
$300\ \rm fb^{-1}$ & $3.3-4.8 \ \sigma$ & 0.23-0.31 &0.12-0.15 \\
$400\ \rm fb^{-1}$ & $3.9-5.5 \ \sigma$ & 0.20-0.27 &0.1-0.14 \\
$500\ \rm fb^{-1}$ & $4.3-6.2 \ \sigma$ & 0.18-0.25 &0.09-0.13 \\
$1000\ \rm fb^{-1}$ &$6.0-8.7 \ \sigma$ & 0.14-0.18 &0.07-0.09 \\
\hline 
\hline
\end{tabular}
\end{table}

\section{Outlook}

By 2007 \babar \ expects to record an integrated luminosity of $\sim 500~
\rm fb^{-1}$. This sample will be sufficient to measure the
$B \rightarrow K^{(*)} \ell^+ \ell^-$ and $B \rightarrow \rho(\omega) \gamma$ 
branching fractions with reasonable precision.
Due to the theoretical uncertainties, however, tests of the Standard Model 
will be rather limited. For example, at a luminosity of
$1~ \rm ab^{-1}$ the presently quoted
theoretical uncertainties for $| V_{td}/V_{ts}|$
will be larger than the extrapolated experimental 
errors.

Observables that are barely affected by theoretical uncertainties and, 
therefore, provide excellent tests of the Standard Model are 
the lepton forward-backward asymmetry as a function of $m_{\ell\ell}$ measured 
in $B \rightarrow K^{(*)} \ell^+ \ell^-$ modes and
direct \CP violation in $B \rightarrow \rho \gamma$ channels. In SM
the lepton forward-backward asymmetry in the $B$ rest frame for dilepton
masses below the $J/\psi$ has a characteristic shape, crossing zero  
at a specific dilepton mass \cite{ali}. 
The zero point is predicted in SM with small 
uncertainties. With limited statistics we will just determine the zero point,
while with high statistics we will be able to
measure the entire distribution. Deviations
from the SM shape will hint to New Physics. In SM \CP asymmetries in 
$B \rightarrow \rho \gamma$ could be as large as $12\%$, but may be modified
considerably in models with minimal flavor violation \cite{ali2}.
Precise measurements of the shape of the lepton forward-backward asymmetry and
direct \CP violation in $B \rightarrow \rho \gamma$, however, require
data samples that are several tens of $\rm ab^{-1}$.

\section{Acknowledgments}
I would like to thank the radiative penguin working group for useful 
discussion and the Norwegian Research Council for support.


\end{document}